\documentclass[aps,prl,superscriptaddress,showpacs,floatfix,twocolumn]{revtex4}

\usepackage{graphicx}   
\usepackage{rotating}
\usepackage{dcolumn}
\usepackage{bm}
\usepackage{epsfig}
\usepackage{amssymb}
\usepackage{multirow}

\begin{document}
\title{Away-side asymmetry of jet correlation relative to reaction plane: a sensitive probe for jet in-medium modifications}
\newcommand{\sunysb}{Department of Chemistry, Stony Brook University, Stony Brook, NY 11794, USA}
\newcommand{\bnl}{Physics Department, Brookhaven National Laboratory, Upton, NY 11796, USA}
\newcommand{\tsukuba}{Institute of Physics, University of Tsukuba, Tenno-dai 1-1-1, Tsukuba, Ibaraki 305-8571, Japan}
\author{Jiangyong Jia}
\affiliation{\sunysb}\affiliation{\bnl}\date{\today}
\author{ShinIchi Esumi}\affiliation{\tsukuba}
\author{Rui Wei}\affiliation{\sunysb}

\begin{abstract}
We proposed a new observable based on two particle azimuth
correlation to study the away-side medium response in mid-central
Au+Au collisions. We argue that a left/right asymmetry may appear
at the away-side by selecting triggers separately in the left and
right side of the reaction plane. A simple model estimation
suggests that the magnitude of such asymmetry could reach 30\% with
details depends on the medium response mechanisms. This asymmetry,
if observed, can help to distinguish competing theoretical models.
\end{abstract}

\maketitle

Recent results from Relativistic Heavy ion Collider (RHIC) indicate
the creation of a new state of matter in Au+Au collisions at
$\sqrt{s_{NN}}$=200 GeV. This matter behaves like a fluid of
strongly interacting quarks and gluons as indicated by its strong
collective flow, low kinetic shear viscosity and large
opacity~\cite{Zajc:2007ey}. It is commonly referred to as the
strongly coupled quark-gluon plasma (sQGP). Current efforts are
focused on the detailed characterization of the properties of the
sQGP.

Energetic jet and back-to-back dijet pairs have been used
extensively to probe the properties of the matter. The interactions
between jet and medium not only lead to a strong suppression of
single hadron yield and away-side dihadron pair yield in central
Au+Au collision at high $p_T$~\cite{Adcox:2001jp,Adler:2002tq},
known as jet-quenching; it also results in characteristic responses
of the medium to the energy deposited by the quenched
jets~\cite{Adams:2005ph,Adler:2005ee}. Such medium responses appear
as several interesting features in the two particle azimuthal angle
correlation at low $p_T$, including the near-side elongation in
pseudo-rapidity (the ridge) and the away-side double-shouldered
structure at
$\pi\pm1.1$~\cite{Adler:2005ee,Adams:2006tj,Adare:2008cq}. The
latter have been interpreted as the possible excitation of conical
flow~\cite{Stoecker:2004qu,CasalderreySolana:2006sq}, however the
exact physics origins are currently under intense
debate~\cite{CasalderreySolana:2007km}.

Our understandings of jet quenching and medium response are based
primarily on how the single particle yield and dihadron correlation
vary with the collision geometry, such as event centrality, system
size and more recently the angle with respect to the reaction plane
(RP). Most theoretical models can describe the centrality and
system size dependence of the single particle and dihadron
data~\cite{Zhang:2007ja,Drees:2003zh}. However they do not do a
good job in describing the RP dependence. The observed anisotropy
for single particle suppression at high $p_T$ seems to be
incompatible with energy loss models~\cite{Bass:2008rv}. The
dihadron correlation also shows a characteristic dependence of the
medium response on the angle with respect to the RP that currently
lacks theoretical descriptions~\cite{Feng:2008an,McCumber:2008id}.
On the experimental side, the physics signals (survived jets and
medium responses) in the previous RP dependence studies are more
susceptible to the influence of hydrodynamic flow than inclusive
studies. This is because the flow modulation is bigger in RP
dependent correlation ($\propto$ $v_{2n}$) than that for the
inclusive correlation ($\propto$
$v_{2n}^2$)~\cite{Bielcikova:2003ku}. In this letter, we propose a
new observable that is sensitive to the medium response mechanisms,
and can offer some insights on the decomposition between jet and
hydrodynamic flow.


\begin{figure}[th]
\epsfig{file=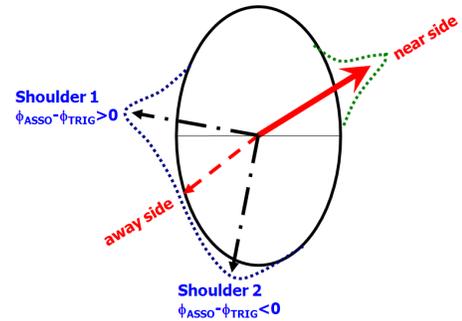,width=0.7\columnwidth}
\caption{\label{fig:1} (Color Online) Schematic view of the path length difference for the two shoulders associated with away-side when trigger is selected
at an angle relative to the reaction plane (indicated by the horizontal line).}
\end{figure}

Our idea is illustrated in Fig.~\ref{fig:1}. A dijet pair is
created and propagated through the medium. One jet exits the medium
and fragments into hadrons containing the trigger; while the
away-side jet loses energy and turns into two branches of medium
responses at angle $\pi\pm1.1$ relative to the trigger. These two
branches have different path length in the medium, and this
difference reaches maximum when trigger angle $\phi_s$ is selected
at $\phi_s\approx\pm\pi/4$. If the magnitude of medium response
depends on the path length, one expects to see a left right
asymmetry. In the case of conical flow, the medium responses
propagate and attenuate in the medium due to finite viscosity. This
may lead to more observed signal at the side with shorter path
length. On the other hand, if the medium response increases with
the path length, then one expects the opposite trend. This can
happen if the medium is pushed outward by the shower gluons which
may be dominantly radiated at large angles relative to the original
hard-scattered partons~\cite{Polosa:2006hb}. Thus the observation
of asymmetry may help us to distinguish competing medium response
mechanisms.

\begin{figure}[th]
\epsfig{file=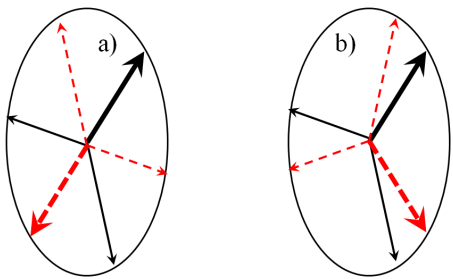,width=0.7\columnwidth}
\caption{\label{fig:2} (Color Online) Two dijets (solid and dashed) that are connected by a) a rotation of $\phi_s\rightarrow\phi_s-\pi$ and b) a change of the sign $\phi_s\rightarrow-\phi_s$. The thick and thin lines indicates the survived near-side jet and medium responses to quenched away-side jet, respectively.}
\end{figure}

But can this asymmetry be observed experimentally? Most RHIC
experiments determines the RP orientation using the second order
event plane which usually has the best resolution. This plane has a
periodicity of $\pi$, which means it does not distinguish two dijet
pairs that are connected by a rotation of $\pi$,
$\phi_s\rightarrow\phi_s-\pi$ (Fig.\ref{fig:2}a). It can, however,
distinguish the two pairs that are connected by a change in the
sign, $\phi_s\rightarrow-\phi_s$
(Fig.\ref{fig:2}b)~\cite{footnote1}. Previous studies did not see
such left right asymmetry with second order event
plane~\cite{Jia:2005ab,Feng:2008an,McCumber:2008id}, because they
fold the trigger angle bins into 0-$\pi/2$ range, which
automatically averages out the asymmetry.

\begin{figure}[th]
\epsfig{file=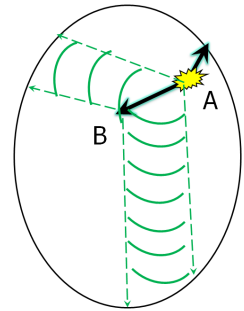,width=0.35\columnwidth}
\caption{\label{fig:3} (Color Online) A more realistic view of the dijet and associated medium responses.}
\end{figure}

Our arguments for left right asymmetry is valid on general grounds,
however a quantitative estimation needs to take into account the
energy deposition, generation and propagation of medium response in
a realistic geometry. Fig.\ref{fig:3} shows a more realistic
scenario, in which the away-side jet is created at point A and is
quenched after traveling to point B. The medium response can be
generated anywhere in between A and B, and propagates along the two
shoulder directions. Due to surface bias of triggered correlation
analysis, point B usually lies deeper into the medium than point A.
For a given medium response mechanism, the magnitude of the
left/right asymmetry is mainly controlled by collision geometry,
i.e. the probability distribution of the hard-scattering point A
and density distribution of the medium. The asymmetry is also
sensitive to the $k_T$ broadening, because it leads to a swing of
away-side jet in azimuthal angle relative to the trigger, which
changes the path length for the medium response.

We estimate the magnitude of the asymmetry using a simple jet
absorption picture based on the Glauber model~\cite{Drees:2003zh}.
In this picture, dijets are generated according to binary collision
density profile in transverse plane with uniform orientation and a
$k_T$ smearing of 0.4 radian according to the $p+p$
data~\cite{Adler:2005ad}. These dijets then traverse the medium
whose density is given by participant density profile. Woods-Saxon
nuclear geometry is used in generating both the collision and the
participant density profiles. For each generated jet at (x,y)
(point A) propagating along direction $(n_x,n_y)$, we calculate the
quenching point B according to survival probability $e^{-\kappa
I}$, where the matter integral $I$ is
\begin{eqnarray}
\nonumber
I&=&\int_{0}^\infty dl\hspace{1mm} l\frac{c\tau}{l+c\tau} \rho{\left(x+\left(l+c\tau\right)
n_x,y+\left(l+c\tau\right)n_y\right)}\\
&\approx& c\tau \int_{0}^\infty dl\hspace{1mm}\rho
\label{eq:2}
\end{eqnarray}
which corresponds to a quadratic dependence of absorption ($\propto
ldl$) in a longitudinal expanding medium ($\propto
c\tau/(c\tau+l)$) with a formation time of $\tau=0.2fm/c$. $\kappa$
is fixed at 0.7 to reproduce the centrality dependence of
$R_{AA}$~\cite{Drees:2003zh}. We select those dijets where one jet
survives and fragments to the trigger particle and the companion
jet is quenched, leading to the associated particles. The
probability for such dijet is
\begin{eqnarray}
f \propto e^{-\kappa I_{jet_1}} \left(1-e^{-\kappa I_{jet_2}} \right)+
e^{-\kappa I_{jet_2}} \left(1-e^{-\kappa I_{jet_1}} \right).
\end{eqnarray}
The spread of the fragmentation of the survived jet is chosen to be
0.3 radian which essentially fixes the near-side width. If jet is
found quenched, the associated medium responses are generated
uniformly between point A and point B at angle $\pi\pm1.1$ relative
to the trigger with an initial azimuth spread of 0.3 radian. To
enable the left/right asymmetry, we assume the medium response is
attenuated according to Eq.~\ref{eq:2}. This leads to an average
attenuation of $\left<e^{-\kappa c\tau \int_{0}^\infty
dl\hspace{1mm}\rho}\right>\approx e^{-1.3}$ for 30-35\% centrality
Au+Au collision. As pointed out in~\cite{CasalderreySolana:2006sq},
a small viscosity of only a few times the universal lower
bound~\cite{Kovtun:2004de} can lead to such level of attenuation.

\begin{figure}[th]
\epsfig{file=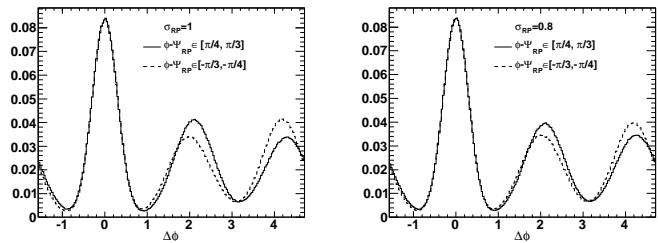,width=1\columnwidth}
\caption{\label{fig:4} The dihadron correlation for triggers angle $\phi$ selected at around $\pm\pi/4$ relative to the reaction plane.
a) is for ideal case and b) is for the case where the reaction plane resolution is 0.8.}
\end{figure}

Fig~\ref{fig:4}a shows the calculated away-side distribution for
trigger hadrons in 30-35\% Au+Au centrality. To maximize the
asymmetry, the angle of the triggers are selected at around
$\phi_s=\pm\pi/4$ relative to the RP. The overall shape of the
away-side and its angular dependence patten is controlled by the
geometry. The multiplicities for the near-side and away-side are
chosen such that the observed amplitudes (average over left and
right) mimic the experimental data. Two features can be clearly
seen. One is the suppression of one shoulder peak relative to the
other, as a result of the path length dependent attenuation of
medium response. The second feature is a small shift of shoulder
peak positions and broadening of their widths, as a result of the
away-side $k_T$ smearing. The overall observed asymmetry is on the
order of 30\%. Fig~\ref{fig:4}b shows the case where the finite RP
resolution is taken into account. The observed asymmetry is rather
sensitive to the accuracy with which one can measure the RP. A
typical resolution of 0.8 used by RHIC experiments reduces the
observed asymmetry by about 50\%.

In non-central Au+Au collision, medium response always coexists
with hydrodynamic flow, which makes the clean separation of the two
effects challenging. The common experimental approach is to assume
that the correlation function consists of a jet-induced signal and
a pure flow term, and to further assume the jet signal is zero at
its minimum $\Delta\phi_{min}$. Recently, this Zero Yield At
Minimum or ZYAM approach~\cite{Adler:2005ee} has been extended to
RP dependence correlation
analysis~\cite{Jia:2005ab,Feng:2008an,McCumber:2008id}. For
triggers selected at a fixed angle $\phi_s$ with ideal RP
resolution, the correlation function defined in \cite{Adare:2008cq}
takes the following form~\cite{Bielcikova:2003ku}:
\begin{eqnarray}
\label{eq:3}
C(\Delta\phi) = {\rm Jet}(\Delta\phi)+\xi(1+2v_2^{\rm assoc}\cos 2(\phi_s+\Delta\phi))
\end{eqnarray}
$v_2^{\rm assoc}$ is the elliptic flow for associated hadrons, and
$\xi$ is the pedestal level in same event measured relative to
mixed event, i.e. $\xi=\langle n_{\rm trig}n_{\rm
assoc}\rangle/\langle n_{\rm trig}\rangle\langle n_{\rm
assoc}\rangle$, where $n_{\rm trig}$ and $n_{\rm assoc}$ denote the
multiplicity of triggers and partners,
respectively~\cite{Adare:2008cq}.  $\xi$ is typically very close to
one in central and mid-central Au+Au
collisions~\cite{Adare:2008cq}, so we simply fix it to unity.

\begin{figure}[th]
\epsfig{file=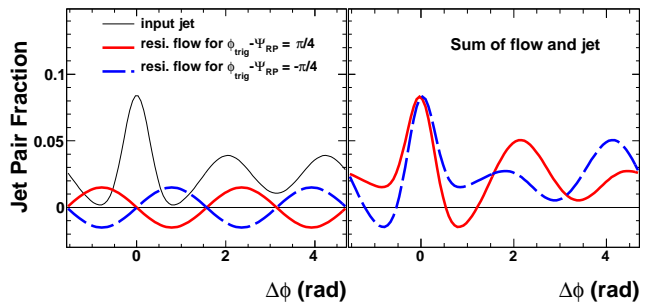,width=1\columnwidth}
\caption{\label{fig:5} (Color Online) Illustrating the effects of the residual flow (assuming to be 5\% of the experimental measured mean value) on the left and right asymmetry. The input jet shape and amplitude (thin line in left panel) is adjusted to mimick the experimentally
measured jet signal (fraction of jet pair over all pairs) for $3-4 \times 1-2$ GeV/$c$ selection from 30-40\% Au+Au collisions \cite{Adare:2008cq}.}
\end{figure}

Equation~\ref{eq:3} indicates that the selection of trigger in a
fixed angle $\phi_s$ leads to a phase shift of $2\phi_s$ in the
flow term. Fig.~\ref{fig:5} illustrates the influence of possible
residual flow left over after flow background subtraction for
triggers selected at $\phi_s=\pm\pi/4$. The input jet shape and
magnitude (thin line in left panel) is adjusted to the
experimentally measured jet signal (jet pair fraction) for $3-4
\times 1-2$ GeV/$c$ bin in 30-40\% Au+Au
collisions~\cite{Adare:2008cq}; the residual flow is assumed to be
the size of the $v_2$ systematic uncertainty, i.e. 5\% of the
measured $v_2^{\rm assoc}$. Fig.~\ref{fig:5} clearly shows that
residual flow can lead to a significant away-side asymmetry at
40-50\% level. This magnitude of the asymmetry is larger than the
real asymmetry shown in the left panel of Fig.~\ref{fig:4}.
However, it is possible that the actual residual flow is
smaller~\cite{shinichi}. Furthermore, residual flow also leads to a
comparable left/right asymmetry at the ZYAM minimum (around $\pm1$
radian), while real medium response of Fig.~\ref{fig:4} does not
show such asymmetry at ZYAM minimum. So by comparing the RP
dependent correlation for trigger selected at $\phi_s$ and
$-\phi_s$, one may be able to detect and constrain possible
residual flow. Such residual flow can be caused by uncertainties in
flow measurements, but it may also indicate that triggered events
have intrinsically different flow from the inclusive events,
possibly due to couplings between jets and medium.

Following the same arguments, it is possible that the near-side
medium response (the ridge) also has left/right asymmetry, caused
by the path length difference between partons emitted at the left
side and those emitted at the right side of the trigger. However,
since the near-side width is quite narrow, the effect due to path
length alone may be too small to be detectable. We leave this for a
future study.

In summary, we propose to study the jet correlation by selecting
triggers separately in the left and right side of the second
harmonic event plane. The correlation signals for triggers selected
this way are sensitive to the left right asymmetry of the geometry
associated with the away-side jet. This asymmetric geometry is
expected to lead to asymmetries for away-side jet shape.
Experimental studies of the shape and magnitude of this asymmetry
may shed light on the underlying medium response mechanisms. Such
studies may also help us to understand possible influences on the
$v_2$ measurements by the coupling between the jets and the flowing
medium. (Indeed both PHENIX and STAR has seen indications of such
asymmetry at the recent Quark Matter conference ~\cite{qm2009}.)

This work is supported by the NSF under award number PHY-0701487
(J.~J, R.~W)  and MEXT and JSPS of Japan (S.~E)


\begin{references}
\bibitem{footnote1} In principle one can distinguish all four cases
    using the $v_1$ event plane, which has a
    periodicity of $2\pi$. On the other hand, the $v_4$ event plane, with a periodicity of
    $\pi/2$, can not resolve the left right asymmetry.
\end{references}
\end{document}